\documentclass[twocolumn,aps,showpacs,preprintnumbers,amsmath,amssymb]{revtex4}

\usepackage{graphicx}% Include figure files
\usepackage{hyperref}% Hyperliens

\begin{document}
\newcommand{\cm}{\ \mbox{cm}^{-1} }

\providecommand{\mean}[1]{\ensuremath{\left \langle #1 \right \rangle}}
\providecommand{\abs}[1]{\ensuremath{\left | #1 \right |}}

\title{Acoustic modes in metallic nanoparticles: atomistic versus elasticity modeling
\\
%Probing the validity of the linear elasticity theory in metallic nanoparticles: a numerical study
}
\author{Nicolas Combe}
\email{combe@cemes.fr}
\affiliation{CNRS ; CEMES (Centre d'Elaboration des Mat\'eriaux et d'Etudes Structurales); BP 94347, 29 rue J. Marvig, F-31055 Toulouse, France.}
\affiliation{Universit\'e de Toulouse ; UPS ; F-31055 Toulouse, France}
\author{Lucien Saviot}
\email{lucien.saviot@u-bourgogne.fr}
\affiliation{Institut Carnot de Bourgogne, Universit\'e de Bourgogne, UMR 5209 CNRS, 9 Avenue A. Savary, BP 47870, F-21078 Dijon Cedex, France }

\providecommand{\ket}[1]{\ensuremath{\left | #1 \right \rangle}}
\providecommand{\bra}[1]{\ensuremath{\left \langle #1 \right |}}
\providecommand{\braket}[2]{\ensuremath{\left \langle #1 | #2 \right \rangle}}
\date{\today}
\def\nbN{\ensuremath{\mathrm{I\!N}}} % IN

\begin{abstract}
The validity of the linear elasticity theory is examined at the
nanometer scale by investigating the vibrational properties of silver
and gold nanoparticles whose diameters range from about 1.5 to 4~nm.
Comparing the vibration modes calculated by elasticity theory and
atomistic simulation based on the Embedded Atom Method, we first show
that the anisotropy of the stiffness tensor in elastic calculation is
essential to ensure a good agreement between elastic and atomistic
models. Second, we illustrate the reduction of the number of vibration
modes due to the diminution of the number of atoms when reducing
the nanoparticles size. Finally, we exhibit a breakdown of the
frequency-spectra scaling of the vibration modes and attribute it
to surface effects. Some critical sizes under which such effects are
expected, depending on the material and the considered vibration modes
are given.
\end{abstract}

%78.67.Bf: optical properties of Nanocrystals and nanoparticles
%63.22.-m: Phonons or vibrational states in low-dimensional structures and nanoscale materials
%63.22.Kn Clusters and nanocrystals
%81.05.Bx Metals, semimetals, and alloys 

\maketitle

\section{Introduction}

Describing structural and elastic properties at the nanometer scale has
recently raised the interest of many scientists. Especially, the
applicability of the linear elastic theory is questioned at these scales.
Various approaches can be used to test the validity of linear
elasticity.  Different groups have tried to measure the elastic
properties, most often the Young modulus, by mechanical deformation
while decreasing the system
size\cite{Chen2006,Wen2008,He2008,Cuenot2004,McDowell2008}.  Another
approach to probe the elastic properties uses acoustic vibration of
nanoparticles (NP) and is mainly based on light or neutron scattering.
Acoustic vibration modes frequencies of NP measured by Raman scattering
or pump-probe experiments are fairly well reproduced by linear elastic
theory calculations even for NPs with sizes of a few
nanometers\cite{Palpant1999,Yadav2006,Yadav2007,Chassaing2009}.

The present work is concerned with the validity of elasticity within the
vibration properties computed for systems whose sizes are smaller than
the ones currently and experimentally explored. To achieve this goal,
we compare the vibration modes calculated by linear elasticity and
atomistic semi-empirical potential calculations in the case of metallic NP
of diameter ranging from 1.4 to about 4~nm. The choice of metallic NP
is motivated by the interest they arouse among the physicists community
due to their plasmon properties opening many promising
applications\cite{Ozbay2006,Maier2007}.  
  
Vibration properties of metallic NPs have been the subject of many
theoretical studies using the linear elasticity or atomistic simulation,
and  experiments.

Since the nineteenth century, the linear elasticity has been used to
calculate the vibrational modes of spherical particles\cite{lamb}: such
calculations generally agree well with experimental measurements of NP
vibrations\cite{Portales2001,Yadav2006}. Different refinements of this
model have been performed taking into account the anisotropy of the
stiffness tensor, the non-spherical shapes of the nanoparticles and matrix effects\cite{Combe2009,Saviot2009,saviot_prl04}. The
improvements of the experimental energy resolution and of the synthesis
of such NPs have allowed to demonstrate subtle effects due to the
anisotropy of the stiffness tensor\cite{Portales2008} and due to the
shape of the NPs\cite{Burgin2008,Chassaing2009}.
 
On the other side, using atomistic simulations, Raman and
Kara\cite{Kara1998,Kara2005} studied the vibrational density of state of
metallic NPs. They showed that there is an enhancement in the
vibrational density of states at low frequencies and an overall shift
of the high frequency band beyond the top of the bulk phonons when
decreasing the NP size. The effect of the capillary pressure induced by the
surface of the NPs has also been exhibited on the vibrational density of
state using  atomistic numerical approaches\cite{Meyer2002,Meyer2003}.
Besides the study of the vibrational density of states, severals
studies have focused on few given vibration eigenmodes of NPs.
Focusing on Raman active vibration modes, the elasticity theory
showed a very good agreement with atomistic calculations in spherical
germanium nanoparticle\cite{Cheng2005,Combe2007}. Recently, a breakdown
of frequency-spectra scaling of respectively, silicon and zinc oxide
nanoparticles have been evidenced  using atomistic simulations and
attributed to surface effects\cite{Ramirez2007,Ramirez2008,Combe2009}.

In this study, we would like to address three issues concerning
the vibration properties of metallic NPs comparing the linear elasticity
predictions and atomistic calculations.

First, we compare the vibration modes of metallic NPs calculated using
either the elasticity theory or an atomistic approach based on an
Embedded Atom Model (EAM). A similar study has been performed by Cheng
et al.\cite{Cheng2005} in the case of germanium NPs. However, their
elastic calculations did not take into account the anisotropy of the
stiffness tensor. In this study, we show that taking into account this
anisotropy in the elastic calculations improves the agreement between
both approaches and that linear elasticity reproduces very well
the vibration modes of NPs having a diameter of a few nanometers.

Second, it is well known, that the vibration properties of a NP
containing $N$ atoms can be described by $3N$ normal vibration modes.
Therefore reducing the size of the NP also reduces the number of normal
vibration modes. Elasticity, as for it, can not take into account this
reduction because of its description using a continuum medium. Some
studies include this reduction in elasticity theory by artificially
limiting the number of modes\cite{tamura83}. In this study, we
illustrate the reduction of the number of normal modes by showing the
disappearance of certain modes while decreasing the NPs size.

Third and finally, while elasticity predicts that the frequency of a
given normal mode scales as the inverse of a characteristic length of
the NP, surface effects can alter this law\cite{Combe2009,Ramirez2007}.
We analyze this scaling in the case of silver and gold spherical NPs and
exhibit a characteristic size under which the frequency scaling
break down. The breakdown of this scaling law is related to the surface
stress and surface relaxation effects. 
  
Section~\ref{sect_method} describes the numerical method used for the
elastic calculations, the atomistic model and the procedure used for
comparing the two models. Section~\ref{sect_result} addresses the three
problems mentioned above. Most of the results presented here concern
silver NPs, except for the last issue for which the case of gold NP has
been considered.
 
\section{Method}
\label{sect_method}
\subsection{Elastic Calculation}
\label{elastic}
\subsubsection{Solving the Navier equation with an anisotropic stiffness tensor}
\label{sect_nav}

The displacement fields $\vec{u}$ associated to each vibration mode
are calculated in the scope of the linear elasticity theory by solving
the Navier equation\cite{landau_elasticity}:

\begin{equation} 
\rho \omega^2 u_i + \sum_{jkl} {\bf C}_{ijkl}\frac{\partial^2 u_k}{\partial x_l \partial x_j}=0 \label{Navier}
\end{equation} 
where $\rho$ is the mass density, $\omega$ the frequency,
$u_i$ the $i^{th}$ component of the displacement fields $\vec{u}$ and
${\bf C}$ the fourth order stiffness tensor. This equation can be
solved analytically for a spherical system in the particular case of an
isotropic stiffness tensor\cite{lamb}. In the case of an anisotropic
stiffness tensor, we turn to a numerical solver. Using the scheme of
Visscher et al.\cite{Visscher1991}, we develop the displacement field
components on a polynomial basis. Solving Eq.~\eqref{Navier} then reduces
to matrix algebra. We use in this study polynomials of order up to 20 to
ensure the convergence of the method. Note that this parameter depends
on the vibration mode of interest: high frequencies vibration modes
commonly correspond to high spatial frequencies which require an extensive
polynomial basis.
%[ J'ai supprim\'e les 2 derni\`eres phrases parce qu'il n'y a pas de modes
%optiques pour Au et Ag. ]

\subsubsection{Modes of interest and identification} 

Severals experimental techniques enable the observation of acoustic
vibration modes, the main ones being inelastic neutron
scattering(INS)\cite{Saviot2008}, Raman scattering, time-resolved
pump-probe experiments and far-infrared absorption. Whereas the absence
of selection rules in INS makes it sensitive to all vibrations, the
three other cited experimental techniques only probe a few acoustic
vibration modes. Therefore, among the very numerous calculated vibration
modes, we are mainly interested in those observable by Raman scattering
experiments. Selection rules for Raman scattering and far-infrared
absorption for an isotropic solid sphere have been given by Duval\cite{Duval}.
Only the breathing (spheroidal, $\ell=0$) and quadrupolar (spheroidal,
$\ell=2$) modes are Raman active. Note that time-resolved pump-probe
experiments only detect the breathing modes of spherical
NP\cite{Arbouet2006}.
%[reference Bachelier Mlayah PRB 2004]

The irreducible representations corresponding to all the vibration modes
have been determined after the displacement fields have been
calculated\cite{Saviot2009}. For a sphere made of a material having a
cubic elasticity (such as silver or gold), the non-degenerate breathing
mode transforms into an A$_{\text{1g}}$ vibration (O$_{\text{h}}$ point
group). The quadrupolar mode splits into E$_{\text{g}}$ and
T$_{\text{2g}}$ vibrations\cite{Portales2008,Saviot2009} and the
corresponding frequency splitting is large making the usage of elastic
anisotropy a key factor for materials such as gold or silver.

\subsection{Atomistic Calculation} 

\subsubsection{Calculation details}

We perform atomistic simulations using the EAM 
potentials developed by Clery and Roseto\cite{Cleri1993}.
Table~\ref{tab1} reports structural and elastic properties calculated in
the scope of this model for silver and gold using the program
GULP\cite{gulp1,gulp2,gulp3}.

\begin{table}
\begin{tabular}{cccc}
\hline
\hline \\
& \mbox{silver } & \mbox{gold} \\
& &\\
\hline \\
a & 4.078 \AA{} & 4.079 \AA{}\\
$C_{11}$ & 132.81 (131) & 187.38 (187)\\
$C_{12}$ & 97.47 (97) & 154.40 (155)\\
$C_{44}$ & 51.11 (51) & 44.71 (45)\\
\hline
\hline
\end{tabular}
\caption{Cell parameter and elastic coefficients of Ag and Au calculated
using the EAM parameters of Cleri and Rosato\cite{Cleri1993}. Elastic
coefficients are expressed in GPa. Experimental data\cite{Simmons1971}
for elastic coefficients are reported in parenthesis.}
\label{tab1}
\end{table}

The calculated elastic coefficients have been used as input in the Navier
equation~\eqref{Navier} in Sect.~\ref{elastic} in order to focus on the
comparison of the models and eliminate possible differences arising from
slightly different elastic tensors.

Spherical NPs are designed with atoms initially placed on a perfect
cubic close-packed crystal structure using the bulk cell parameters of
the modeled material. Our NPs do not specifically reproduced NPs with
magic number\cite{Martin1996}. In addition, our construction procedure
inevitably produces some steps and facets especially for very small NP.
The total energy is then relaxed using a conjugate gradient algorithm:
this procedure ensures that all vibration eigenfrequencies are real.
Our model omits the possible existence of surface reconstructions and
dangling bonds: these aspects could correctly be addressed using more
sophisticated techniques (for instance DFT or tight binding). The
dynamical matrix is then computed and diagonalized to obtain
eigenfrequencies and eigenvectors. Eigenvectors are normalized for the
usual scalar product in the real space of dimension $3N$ with $N$ the
number of atoms in the NP. In the following, we study spherical NPs from
19 to 2072 atoms corresponding to diameter ranging from 1.6 nm to 4.3
nm for silver NPs and from 10 to 2491 atoms corresponding to diameter
ranging from 1.4 nm to 4.5 nm for gold NPs. The validity of the EAM
potential for a 10 atoms NP is arguable and we admit that an {\it
ab-initio} density functional calculation would be more appropriated in
this case. However, in the following we compare the vibration
properties of NPs on a large range of sizes. As a result, it does not
seem judicious to use different atomistic models for different sizes
since it would also result in a slight modification of the elastic
stiffness tensor. As a matter of fact, though quantitative results for
the smallest NPs would need a more accurate modeling, we believe that
results described below are semi-quantitative in the sense that they
give a good sketch of the vibration properties of NPs while reducing
size.
 
In the following, we will refer to atomistic modes for vibration modes
calculated in the framework of the atomistic EAM calculations.

\subsubsection{Modes identification}

While the linear elasticity theory assumes a continuous medium
and thus provide a displacement field for each vibration mode,
atomistic vibration modes are described by the data of each atom
moves\cite{Combe2007}. For this reason, and because the precise symmetry
of the NPs shapes is not fixed when relaxing the energy, the symmetry
analysis we did in the case of the elastic calculation becomes more
troublesome in the case of atomistic modes. Instead, we prefer to
project elastic modes on atomistic ones as it has already been done in
the case of semi-conductor NPs\cite{Cheng2005}. The scalar product of
the $p$th elastic mode with the $q$th atomistic mode reads:
\begin{equation}
< \vec{u}^{elas}_p |\vec{u}^{atomic}_q> = \sum_{i \in atoms}
\vec{u}^{elas}_{p}(\vec{R}_i) . \vec{u}^{atomic}_{q, i}
\label{eq_scalar}
\end{equation}
where $\vec{R}_i$ are the atomic
positions in the atomistic model and $\vec{u}^{atomic}_{q, i}$ is the
displacement of atoms $i$ located at position $\vec{R}_i$. Note that
atomistic modes are normalized for this scalar product. We also
normalize elastic modes so that:

\begin{equation}
  \sum _ q < \vec{u}^{elas}_p |\vec{u}^{atomic}_q>^2 =1
\label{normalisation}
\end{equation}

%[ Est-ce que cette op\'eration est ''violente''? Les facteurs de
%renormalisation sont-ils tous \`a peu pr\`es \'egaux? ]

A more relevant quantity is the sum of such squared projections of elastic modes
onto atomistic ones for all degenerate elastic modes having the same
irreducible representation and eigenfrequency\cite{Saviot2009}.
More precisely, if $\Omega_{elas}$ is a set of such elastic modes,  we define its projection on atomistic ones as: 
\begin{equation} 
Q_{q}^{\Omega_{elas}}=\sqrt{\frac{\sum_{p\in\Omega_{elas}}<\vec{u}^{elas}_p|\vec{u}^{atomic}_q>^2}{\mathbb{N}_{\Omega_{elas}}}}\label{set_projection}
\end{equation}
where $\mathbb{N}_{\Omega_{elas}}$ is the number of modes in the set  $\Omega_{elas}$. The denominator in Eq.\eqref{set_projection} ensures the normalisation condition (similar to Eq.\eqref{normalisation}) for the projection of a set of elastic modes. 
 The quantity $Q_{q}^{\Omega_{elas}}$ does not depend on the choice of degenerate elastic
displacements. As such it is suitable for the comparison with atomistic
calculations and will be used in the following.

 \section{Results}
\label{sect_result}

\subsection{Elastic regime: Relevance of the anisotropy of the stiffness tensor}

\subsubsection{Mode Projection} 

When surface effects are negligible, i.e. the NPs size is larger than
a characteristic size which will be established below, the elasticity
theory reproduces very well the atomistic calculations.

\begin{figure}[h]
\includegraphics[width=\columnwidth]{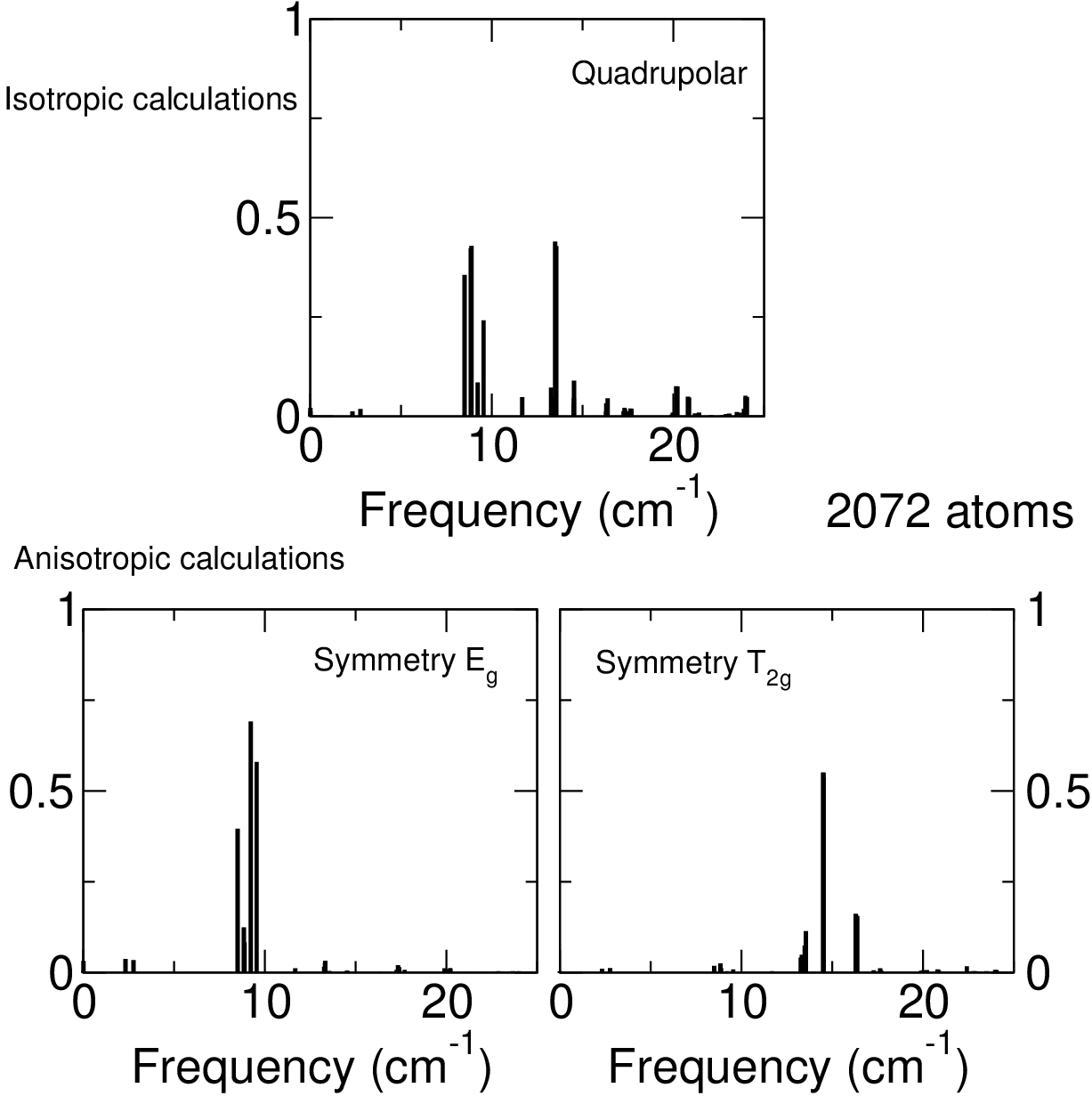}
\caption{(color online) Projection of the fundamental quadrupolar
elastic vibration modes of a 4.3~nm diameter silver spherical
NP, calculated using the linear elasticity theory on atomistic
ones using the scalar product given in Eq.~\eqref{eq_scalar}
and Eq.~\eqref{set_projection} . Results are reported for both
isotropic (above) and anisotropic (below) calculations. For isotropic
calculations, the quadrupolar elastic vibrations modes form a set of
5 degenerated modes with the same irreducible representations. In the
anisotropic calculations, quadrupolar modes splits in two sets of modes
of respective symmetry $E_g$ and $T_{2g}$. Abscissa denotes atomistic
modes using their frequency in $\cm$.
}
\label{fig1}
\end{figure}

Fig.~\ref{fig1} reports the projection of the set of fundamental quadrupolar
$E_g$ and $T_{2g}$ modes calculated by elasticity theory on atomistic
modes for a 4.3~nm diameter silver spherical NP using Eq.~\eqref{set_projection}. In order to show the
relevance of the anisotropy of the stiffness tensor in elastic
calculations, Fig.~\ref{fig1} reports the projection of the set of fundamental
quadrupolar elastic modes calculated using both an isotropic and an
anisotropic stiffness tensor. The anisotropic elastic calculations
use the stiffness tensor calculated from the EAM atomistic model and
reported in Tab.~\ref{tab1}. The isotropic calculation uses a stiffness
tensor obtained using the 3D averaged sound velocities. Atomistic
calculations intrinsically include the anisotropy of the elastic properties. 

In the isotropic elastic approximation the displacements of the 5
quadrupolar vibrations ($m = 0, \pm 1, \pm 2$) correspond to a single
frequency. In the anisotropic elastic approximation, this degeneracy is
partially lifted. As a result, 2 of these ``quadrupolar'' vibrations
have the same frequency (E$_{\text{g}}$ vibrations) and the 3 others
have a different frequency (T$_{\text{2g}}$ vibrations). It should be
noted that the main effect of elastic anisotropy is to split the
frequencies into 2 groups but that the corresponding displacements are
almost unaffected\cite{Saviot2009}. Fig.~\ref{fig1} provides a simple
way to check that taking into account anisotropy in the elastic
calculations significantly improves the description. The quadrupolar
modes from the isotropic elastic calculation have a significant
projection onto atomistic modes over a much larger range than the
corresponding anisotropic ``quadrupolar'' elastic modes.  Therefore the
anisotropic description is clearly an improvement. The E$_{\text{g}}$
vibrations have a significant projection over a narrow frequency range.
For the T$_{\text{2g}}$ vibrations, the projections are significant over
a larger frequency range, but this is probably due to the light mixing
which occur in the elastic calculations between the T$_{\text{2g}}$
branches coming from the isotropic quadrupolar and torsional ($\ell=3$)
modes\cite{Saviot2009}. This mixing is different within the atomistic
approach which explains the appearance of an additional peak at higher
frequency ($\sim 16.3 \cm$).

In the following, elastic calculations always use the anisotropic
stiffness tensor defined from Tab.~\ref{tab1}. 
 
\subsection{Size dependence and mode number reduction}
\label{number_reduction}

\begin{figure}[h]
\includegraphics[width=\columnwidth]{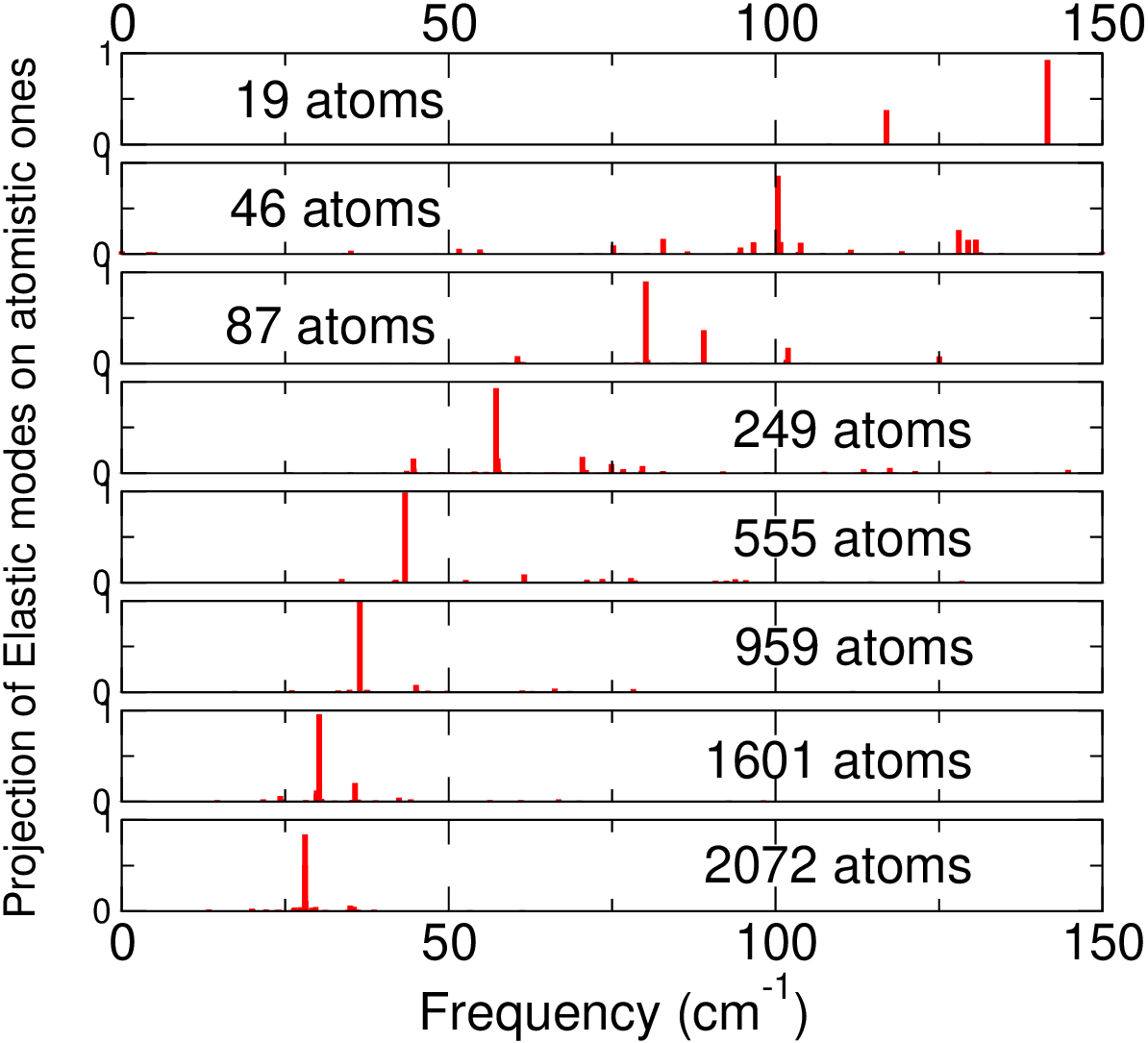}
\caption{(color online) Projection of the breathing vibration modes of silver spherical
NP calculated using linear elasticity theory onto the atomistic modes using
the scalar product given in Eq.~\eqref{eq_scalar}. The number of atoms
in the NP is reported on each graph. The frequency in $\cm$ of the
atomistic modes is used on the abscissa.}
\label{fig2}
\end{figure}

Following the work of Cheng et al.\cite{Cheng2005} we report in
Fig.~\ref{fig2} the evolution of the projection of the fundamental
breathing mode for different silver spherical NP sizes. This particular
mode is the A$_{\text{1g}}$ vibration with a frequency
close to the isotropic breathing mode and having the largest volume
variation\cite{Saviot2009}.
The projection of this elastic mode onto the atomistic ones is
all the more peaked on a single atomic mode for big NPs. This
observation agrees with the expectation that elasticity works well for
large NPs for which the continuous medium approximation is more appropriated. Note that
even for a very small NP, elasticity reproduces fairly well the
atomistic modes. The same conclusion has been raised by Cheng et
al.\cite{Cheng2005} while studying spherical germanium NPs.

\begin{figure}[h]
 \includegraphics[width=\columnwidth]{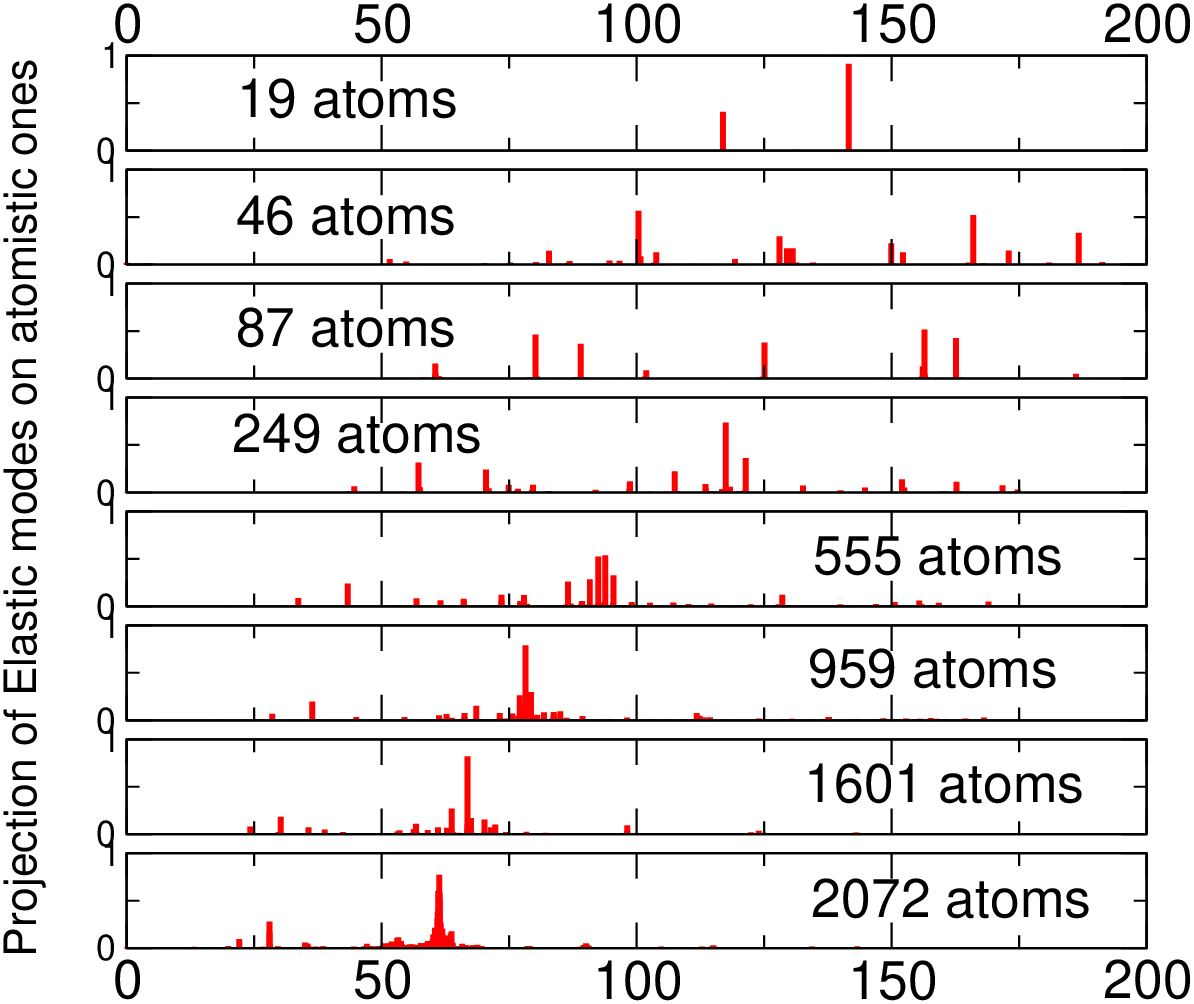}
 \caption{(color online) Projection of the first overtone of the breathing mode of
silver spherical NPs calculated using linear elasticity theory onto the
atomistic modes using the scalar product given in Eq.~\eqref{eq_scalar}.
The number of atoms in the NP is reported on each graph. The frequency
in $\cm$ of the atomistic modes is used on the abscissa.}
\label{fig3}
\end{figure}

Reducing the size of the NP diminishes its number of atoms. Thus, it
also reduces the number of vibration modes. From Fig.~\ref{fig2}, the
breathing mode is evidenced even for the very small NP containing only
19 atoms. However, we expect that higher harmonics may disappear from
the atomistic spectra while decreasing the NP size. In Fig.~\ref{fig3},
we report the projection of the first overtone of the breathing mode
(A$_{\text{1g}}$ vibration with a frequency close to the first overtone
of the isotropic breathing mode and having the largest volume variation)
of silver spherical NPs on atomistic modes as a function of the NP
size.

Contrarily to the fundamental breathing mode, the projection of the first
overtone spreads over several atomistic modes for NPs made of 249 atoms
and less. It is still possible to assign a main corresponding atomistic
mode for a 249 atoms NP but no atomistic modes seem to describe the
elastic mode for the 87 atoms NP. Instead, this mode projects onto
several atomistic modes with a weak scalar product. For smaller NPs,
it mainly projects onto
the same atomistic mode as the fundamental breathing mode. This behavior is a
manifestation of the disappearance of higher frequency elastic
vibration modes due to the decrease of the number of atoms in the NPs.
Indeed, while approximately 70 elastic modes have a frequency smaller
than that of the fundamental breathing mode, this number increases
to approximately 500 for its first overtone. These numbers are to
be compared with $3N-3$ which is the number of vibrations for a
nanoparticle made of $N$ atoms. We can crudely estimate that the
first overtone of the breathing mode is not well defined for $N$ such
that $3N < 500$ which is in agreement with the results presented in
Fig.~\ref{fig3}.
The projection onto the same atomistic mode as the fundamental breathing
mode is due to the very close symmetry of the two modes. Note that in
Fig.~\ref{fig3}, the abscissa scale extends over the whole spectrum of
vibration states of the atomistic model. Tamura and
Ichinokawa\cite{tamura83} defined some maximum frequencies for spheroidal and
torsional modes depending on the number of atoms in the NP in order to
use the elastic model to calculate the specific heat of small NPs.
A similar rule might apply to anisotropic calculations. But checking the rules proposed by Tamura and Ichnokawa from our
calculations is beyond the scope of this paper.
Current computer facilities make the calculations of vibration modes of small NPs easy so that the rules
proposed by Tamura and Ichonakawa is not crucial nowadays for
interpreting experimental data.	 
  
\subsection{Size dependence: breakdown of elasticity prediction}

\subsubsection{Power Law}

\begin{figure}[h]
%\centerline{\epsfxsize=8cm \epsfbox{quad0_breath01_AG} }
\includegraphics[width=\columnwidth]{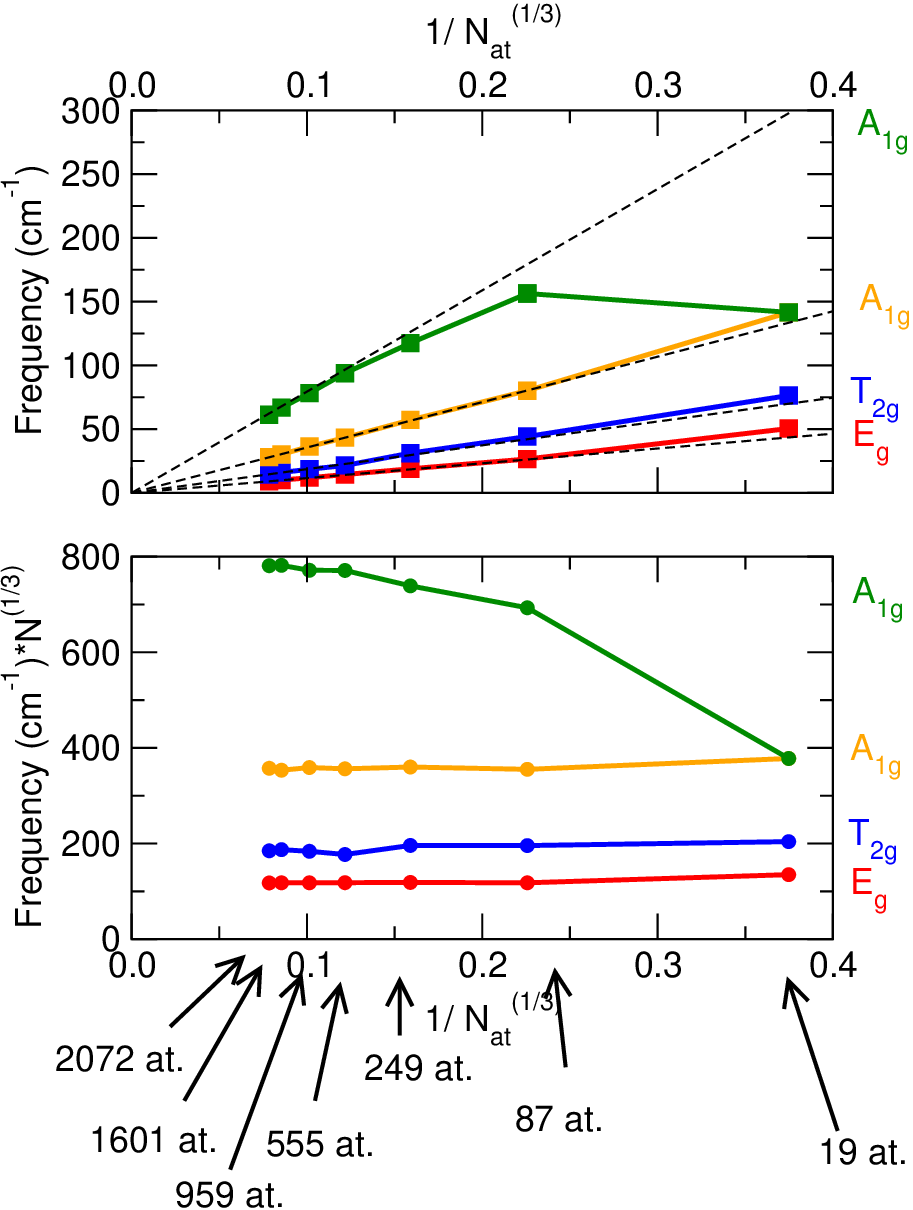}
\caption{(color online)(above) Frequencies of the atomistic modes onto which the
E$_{\text{g}}$ and T$_{\text{2g}}$
modes coming from the fundamental quadrupolar modes and the
A$_{\text{1g}}$ modes coming from the fundamental (low frequency) and first overtone (high frequency)  of the
breathing mode preferentially project in spherical silver
NPs as a function of the inverse of the number of atom at the power one
third. Dashed lines correspond to frequencies calculated by the elasticity theory: the number of atoms is calculated from the ratio of the NP volume on the atomic volume computed from atomistic calculations.
(below) Product of the frequency and the number of atoms at the power
one third as a function of the inverse of the number of atoms at the
power one third. Linear elasticity predicts horizontal straight lines.}
\label{fig4}
\end{figure}

\begin{figure}[h]
%\centerline{\epsfxsize=8cm \epsfbox{quad0_breath01_AG} }
\includegraphics[width=\columnwidth]{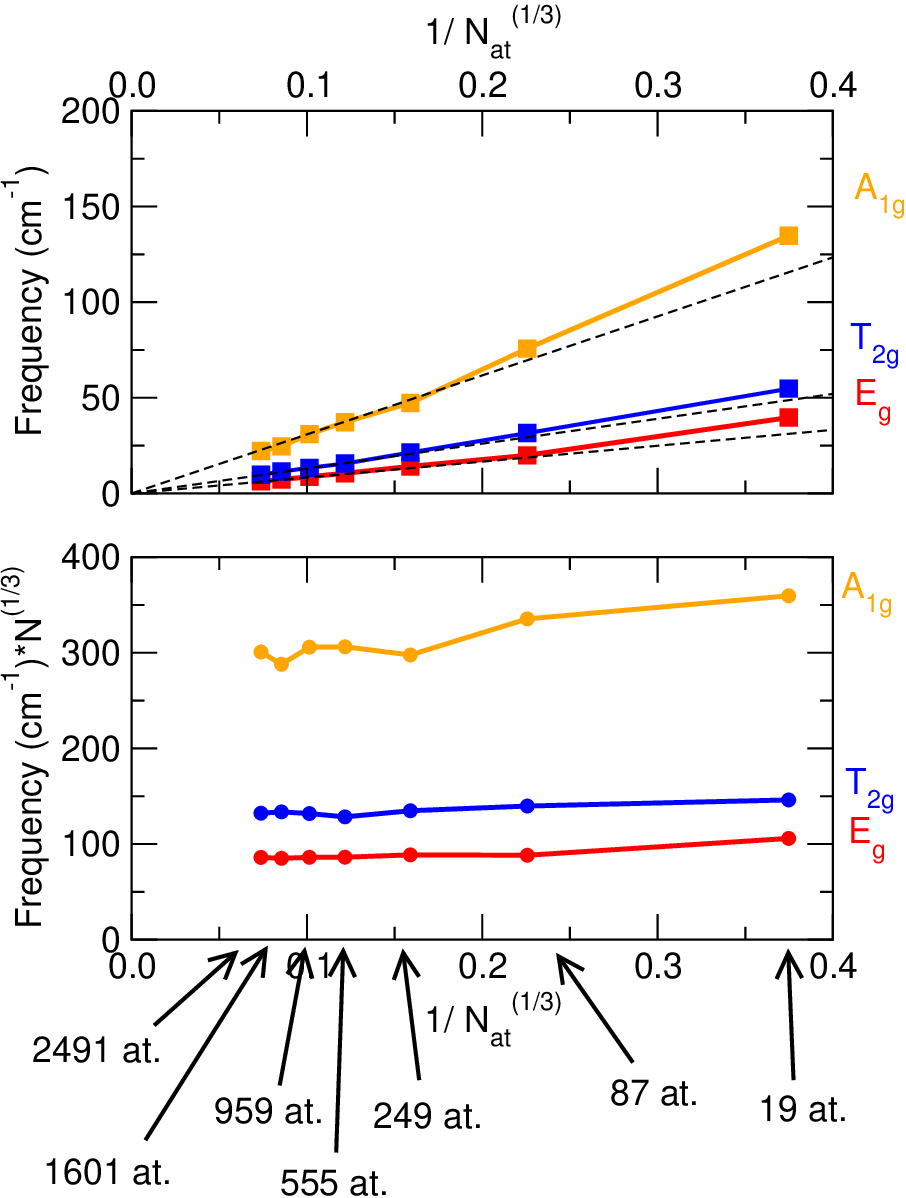}
\caption{(color online) (above) Frequencies of the atomistic modes onto which the
E$_{\text{g}}$ and T$_{\text{2g}}$
modes coming from the fundamental quadrupolar modes and the
A$_{\text{1g}}$ mode coming from the fundamental
breathing mode preferentially project in spherical gold
NPs as a function of the inverse of the number of atom at the power one
third. Dashed lines correspond to frequencies calculated by the elasticity theory: the number of atoms is calculated from the ratio of the NP volume on the atomic volume computed from atomistic calculations.
(below) Product of the frequency and the number of atoms at the power
one third as a function of the inverse of the number of atoms at the
power one third. Linear elasticity predicts horizontal straight lines.}
\label{fig5}
\end{figure}

Using linear elasticity theory and from Eq.~\eqref{Navier}, the
frequency of a given vibration mode of a solid sphere scales as $1/R$
where $R$ is the radius of the sphere. This law which we will refer to
as the Elastic Frequency Law (EFL) in the following, can be used to probe
the validity of the elasticity theory. Recently, the validity of the EFL
has been questioned for very small NP for which surface effects become
non negligible\cite{Combe2009,Ramirez2007}.

In this section, we investigate the validity of the EFL in the case of
metallic NP. We thus report the frequency of a given atomistic mode as a
function of $1/N^{1/3}$(proportional to the radius of the NP) with N the
number of atoms in the NP. This atomistic mode is defined as the one on
which the projection of elastic modes is maximum ignoring the relative
values of this projection compared to the other modes. 
Figure~\ref{fig4} reports the frequencies of four selected atomistic
modes for spherical silver NP as a function of $1/N^{1/3}$. We first
choose four elastic modes: one E$_{\text{g}}$ mode and one T$_{\text{2g}}$
mode coming from the fundamental isotropic quadrupolar mode and two
A$_{\text{1g}}$ modes coming from the fundamental and first overtone of
the isotropic breathing mode. Then for each elastic mode and number of
atoms $N$, we consider only the atomistic mode onto which the elastic
mode has the largest projection and report its frequency as a function of $1/N^{1/3}$. To identify the possible breakdown of
the scaling law, we also plot the product of the frequencies and
$N^{1/3}$ as a function of $1/N^{1/3}$.
Note that the evaluation of the radius of a NP in the atomistic model may involve some
technical issues regarding the precise definition of the surface,
especially for small NPs: the surface is not exactly spherical and the
position of the surface can be slightly different from the position of
the surface atoms when taking into account atomistic radius.  As
a consequence, we prefer to report frequencies as a function of the unambiguous quantity 
$1/N^{1/3}$. \\

 Figure~\ref{fig4} shows that
atomistic calculations agree well with the EFL for the atomistic
quadrupolar modes and the atomistic fundamental breathing mode. Except
for the smaller NP with 19 atoms, for which the frequency differs
slightly from the EFL, all the other points follow the EFL. The
frequencies of the atomistic fundamental breathing mode and the two
fundamental quadrupolar modes differ by about $6\%$ ($8 \cm$) and $5\cm$
and $6\cm$ respectively for 19 atoms NP. Difference for bigger NP are
smaller than $1\cm$.

Concerning the first overtone of the breathing mode,
Fig.~\ref{fig4} shows that the atomistic frequency starts to differ
significantly from the EFL for less than about 250 atoms corresponding
to a characteristic diameter of 2.2~nm. This breakdown of the EFL could
be related to surface effects\cite{Combe2009}(we exclude at this point the case of the 19 atoms NP). Surface and surface
relaxation effects thus do not affect equivalently the different
vibration modes. The relaxation of the surfaces spreads on a
characteristic size $\lambda$ of a few Angstroms under the
surface\cite{Meyer2002}. We can reasonably expect that modes with a
small wavelength (such as harmonics of isotropic modes) are more
affected than longer wavelength ones (or fundamental mode) by the
presence of the surface. Obviously this claim also depends
on the relative volume affected by the surface relaxation compared to the
unaffected volume in the NP: frequencies of optical modes or border zone
phonon modes in a semi-infinite medium are not affected by the presence
of the surface. Finally, for the 19 atoms NP, the atomistic first
overtone of the breathing mode is identical to the atomistic fundamental
breathing mode because of the reduction of the number of modes and
symmetry arguments as already discussed in Sect.~\ref{number_reduction}.
 
In addition, we performed the same study for gold NP to investigate the
dependence of our results on the material. Figure~\ref{fig5} reports the
frequency of the fundamental quadrupolar and breathing atomistic modes
for a spherical gold NP as a function of $1/N^{1/3}$. While the
atomistic quadrupolar modes reproduce the EFL fairly well, the atomistic
fundamental breathing mode significantly shifts from the EFL at a
characteristic size of about 87 atoms corresponding to a diameter of
about 1.6~nm: the frequency of the atomistic fundamental breathing mode
differs by about $18\%$ ($21 \cm$) from the EFL for 19 atoms NP and by
about $12\%$ ($8 \cm$) for the 87 atoms NP. As in the case of silver NP,
we attribute the breakdown of the EFL to surface relaxations effects.
Note that our results suggest that surface effects seems to be more
important in gold NPs than in silver NPs (for which the atomistic
fundamental breathing mode reproduce the EFL fairly well except for the
smallest studied NP). This result is corroborated by the fact that
surface relaxation in gold NP extends over a larger range than in
silver\cite{Meyer2002, Sun2001}, and that the surface stress (to which surface relaxation is related) in gold is about two times the one in silver: this last result has been obtained  for different surface orientations by atomistic simulations based on the EAM~\cite{Gumbsch1991,Ackland1987} and on the Modified EAM~\cite{Wan1999}.

In addition, to rule out the dependence of our results on the precise
shape of the NP, we have also performed simulations on truncated
cuboctahedron silver NPs: no significant deviation from the results
presented here has been observed.

\section {Conclusion} 

By comparing atomistic and linear elasticity calculations, we have
exhibited the importance of the anisotropy of the stiffness tensor in
elastic calculations. Especially, both atomistic and anisotropic elastic
calculations lift the degeneracy of the quadrupolar modes.
By projecting the elastic modes onto the atomistic ones, we evidenced the good
agreement between the two models for the fundamental breathing mode even
for very small NPs. These two results suggest the applicability of the
linear elastic theory to less than 2 nm diameter silver NPs. However, we
also evidenced the decreasing number of vibration modes and the breakdown
of the EFL when reducing the size.
We conclude, that the applicability of the linear elasticity depends on
the considered vibration mode and the NP size. Our results suggest that
the breakdown of elasticity occurs for bigger NPs for high harmonics
rather than
for fundamentals modes. Moreover, critical sizes depend on the material.
Such breakdown to our knowledge, has never been evidence experimentally
in any material because of the difficulty to synthesize NPs of diameter
of less than 2~nm with a narrow size distribution or to perform
single particle measurements on such objects. Our results suggest that the observation
of the breakdown of the vibration frequency scaling in silver NPs will
be hazardous, since it would imply some experiments on silver NP (or
cluster) of about 20 atoms for the fundamental breathing and quadrupolar
modes. This task though difficult, seems more reachable in the case of
gold NP since it would require measurements on NPs with less than about 100 atoms.
Finally, because the shape of very small NPs can hardly be described by
a sphere, we have confirmed through simulations on truncated cuboctahedron
silver NPs that our results do not strongly depend on the exact shape of
the NPs.

\begin{acknowledgments}
NC thanks the support of the CNano GSO program and the PPF Grand Sud-Ouest. 
\end{acknowledgments}

%\bibliographystyle{apsrev}
%\bibliography{/home/combe/ARTICLE/PUBLICATIONS/biblio}

\end{document}